\documentclass[review,sort&compress]{elsarticle}

\usepackage{graphicx}
\usepackage{epstopdf}
\usepackage{amssymb}
\usepackage{latexsym}

\journal{Journal of Sound and Vibration}

\begin{document}

\begin{frontmatter}

\title{Universal response of optimal granular damping devices}

\author[1,2]{Mart\'{\i}n S\'anchez}\ead{sanchez.martin@frlp.utn.edu.ar}
\author[3]{Gustavo Rosenthal}\ead{guser@frlp.utn.edu.ar}
\author[1]{Luis A. Pugnaloni\corref{cor}}\ead{luis@iflysib.unlp.edu.ar}
\cortext[cor]{Corresponding author. Tel.: +54-221-423-3283; fax: +54-221-425-7317}

\address[1]{Instituto de F\'{\i}sica de L\'{\i}quidos y Sistemas Biol\'{o}gicos (CONICET La Plata, UNLP), Calle 59 Nro 789, 1900 La Plata, Argentina.}
\address[2]{Centro de Ensayos Estructurales, Facultad Regional Delta, Universidad Tecnol\'ogica Nacional, Av. San Mart\'{\i}n 1171, B2804GBW Campana, Argentina.}
\address[3]{Departamento de Ingenier\'{\i}a Mec\'anica, Facultad Regional La Plata, Universidad Tecnol\'ogica Nacional, 60 esq. 124 S/N, 1900 La Plata, Argentina.}

\begin{abstract}
Granular damping devices constitute an emerging technology for the attenuation of vibrations based on the dissipative nature of particle collisions. We show that the performance of such devices is independent of the material properties of the particles for working conditions where damping is optimal. Even the suppression of a dissipation mode (collisional or frictional) is unable to alter the response. We explain this phenomenon in terms of the inelastic collapse of granular materials. These findings provide a crucial standpoint for the design of such devices in order to achieve the desired low maintenance feature that makes particle dampers particularly suitable to harsh environments.
\end{abstract}


\end{frontmatter}

\section{Introduction}

\begin{figure}[t]
\begin{center}
\includegraphics[width=0.6\columnwidth]{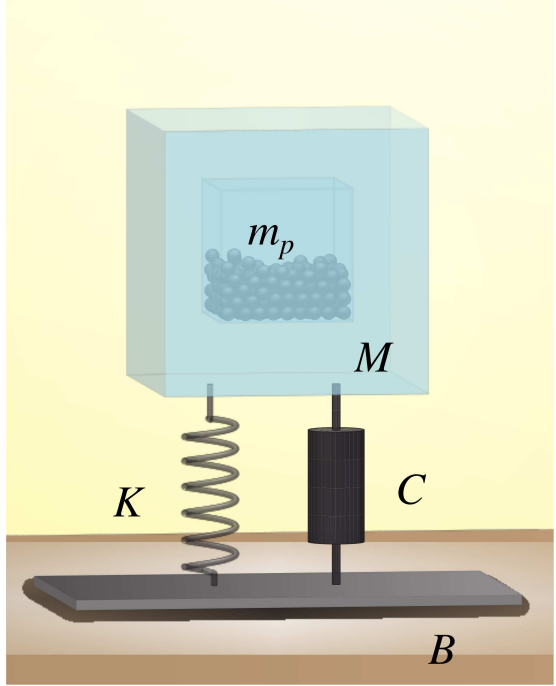}
\end{center}
\caption {(Color online) Schematic view of a particle damper. ($M$) the primary mass of the structure, ($m_p$) the total mass of the particles in the enclosure, ($K$) the spring constant, ($C$) the viscous structural damping, ($B$) the vibrating base where the displacement is imposed.}
\label{fig1}
\end{figure}

Granular dampers (or particle dampers, PDs) are devices aimed at the attenuation of mechanical vibrations by exploiting the dissipative character of the interaction between macroscopic particles. A PD consists in a number of particles (grains) enclosed in a receptacle that is attached or embedded in a vibrating structure (see Fig. \ref{fig1}). The motion of the grains inside the enclosure, as the structure vibrates, is able to dissipate part of the energy through the non-conservative collisions, so reducing the vibration amplitude. This emerging technology can replace the widely used viscous  and viscoelastic dampers in particular applications where extreme temperatures (either low or high) are involved or where low maintenance is required. The leading sector in this regard is the aerospace industry \cite{Simonian, Panossian, Ehrgott}. However, the automotive \cite{Xia} and oil and gas \cite{Velichkovich} industries are catching up in recent years. PDs are the descendant of the older impact dampers designed to operate by the use of a single body inside an enclosure \cite{Lieber, Duncan}. PDs are now preferred over impact dampers due to the lower noise level they produce. The performance of a PD depends on a number of design characteristics such as the relative size and shape of the particles and the enclosure, the total weight $m_p$ of the particles, the number $N$ of grains, the working vibration intensity and frequency, etc. These have been studied to some extent in the last two decades \cite{Friend, Saeki, Liu}. However, less attention has been paid to the role that the grain-grain interaction plays in these systems. 

In this paper, we show that the response of a PD is independent of the particle-particle interaction to the extent that even friction or inelastic collisions can be suppressed without altering the vibration attenuation. This effect is explained in terms of the effective zero restitution of the granular system caused by the effective \emph{inelastic collapse} \cite{Kadanoff}. The inelastic collapse in dense granular materials refers to the effect by which the system can dissipate its entire kinetic energy in a short finite time even if collisions have a very high restitution coefficient. In dense systems, the number of collisions grows so rapidly that even a minute dissipation in each collision suffices to make the system as a whole fully dissipative. We show that this interpretation allows us to set the limits to the universal response. Crucial implications for the design and maintenance of PDs are discussed.

\section{DEM Simulations}

We have carried out simulations of a PD by means of a Discrete Element Method (DEM). A number N (with $5<N<250$) of spherical particles of mass $m=9.08 \times 10^{-4}$ kg are deposited in a prismatic enclosure (see Fig. \ref{fig1}) of mass $M=2.37$ kg, base $L \times L$ ($L=0.03675$ m) and height $L_z$ (with $0.057$ m$<L_z<0.282$ m). The enclosure is attached to a vibrating base by means of a Hookean spring (spring constant $K=21500$ Nm$^{-1}$) and a viscous damper of low dissipation constant ($C=7.6$ Nsm$^{-1}$). The particles interact through a Hertz-Kuwabara-Kono \cite{Brilliantov} normal contact force ($F_\mathrm{n} = -k_\mathrm{n}\alpha^{3/2}-\gamma_\mathrm{n}\upsilon_\mathrm{n}\sqrt{\alpha}$) plus a tangential force ($F_\mathrm{t} = -\min\left(\left|\gamma_\mathrm{t}\upsilon_\mathrm{t}\sqrt{\alpha}\right|,\left|\mu_\mathrm{d}F_\mathrm{n}\right|\right)\rm{sgn}\left(\upsilon_\mathrm{t}\right)$) that implements the frictional property of the grain surfaces \cite{Schafer}. Here, $\upsilon_\mathrm{n}$ and $\upsilon_\mathrm{t}$ are the relative normal and tangential velocities and $\alpha=r_{ij}-d$ the virtual overlap between the interacting particles $i$ and $j$. The diameter of the particles is $d=0.006$m. The parameters that set the grain-grain interactions are: $k_n=4.02 \times 10^9$ Nm$^{-3/2}$ (which corresponds to the Young modulus $E=2.03 \times 10^{11}$ Nm$^{-2}$ and Poison ratio $\nu=0.28$ for steel), $0< \gamma_n< 1 \times 10^4$ kgm$^{-1/2}$s$^{-1}$, $0<\gamma_t< 1 \times 10^4$ kgm$^{-1/2}$s$^{-1}$ and $\mu_\mathrm{d}=0.3$. The grain-wall interaction is taken equal to the grain-grain interaction. The vibrating base and the enclosure are constrained to move only in the vertical $z$-direction, whereas the particles can move in the three-dimensional volume of the receptacle subjected to the action of gravity. The motion $z_{base}(t)$ of the base is set to a harmonic function [$z_{base}(t)=0.0045\cos(\omega t)$ m] whose frequency $\omega = 2\pi f$ is varied in the range $0.5 \mathrm{Hz}< f <30 \mathrm{Hz}$. We monitor the amplitude of the oscillation $z_{max}$ of the enclosure in response to the base vibration, the motion of the grains inside and the energy dissipated in each cycle. All analysis is done over the stationary state of the system for each given $f$ after any transient has died out.

\begin{figure}[t]
\includegraphics[width=0.49\columnwidth]{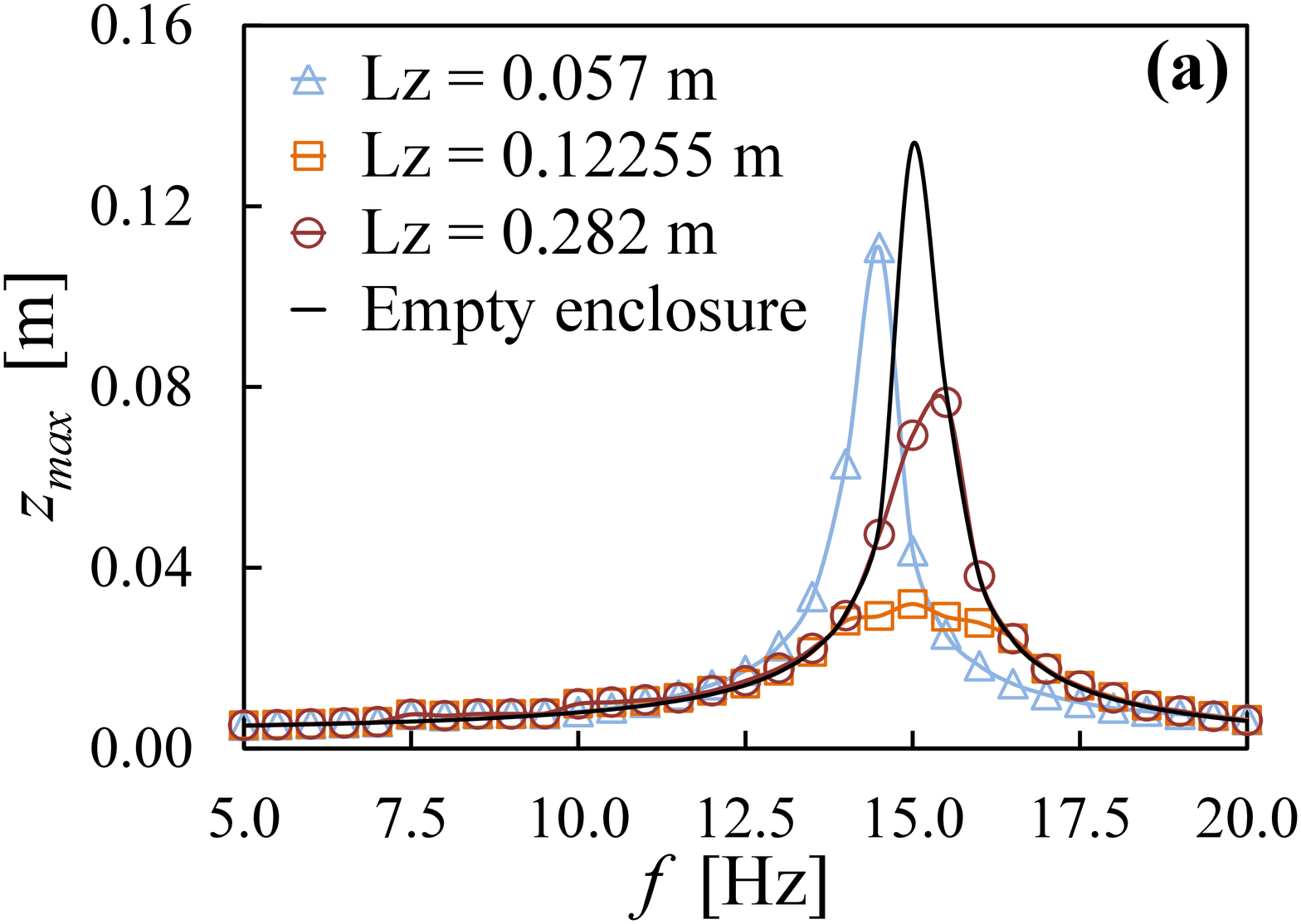}
\includegraphics[width=0.49\columnwidth]{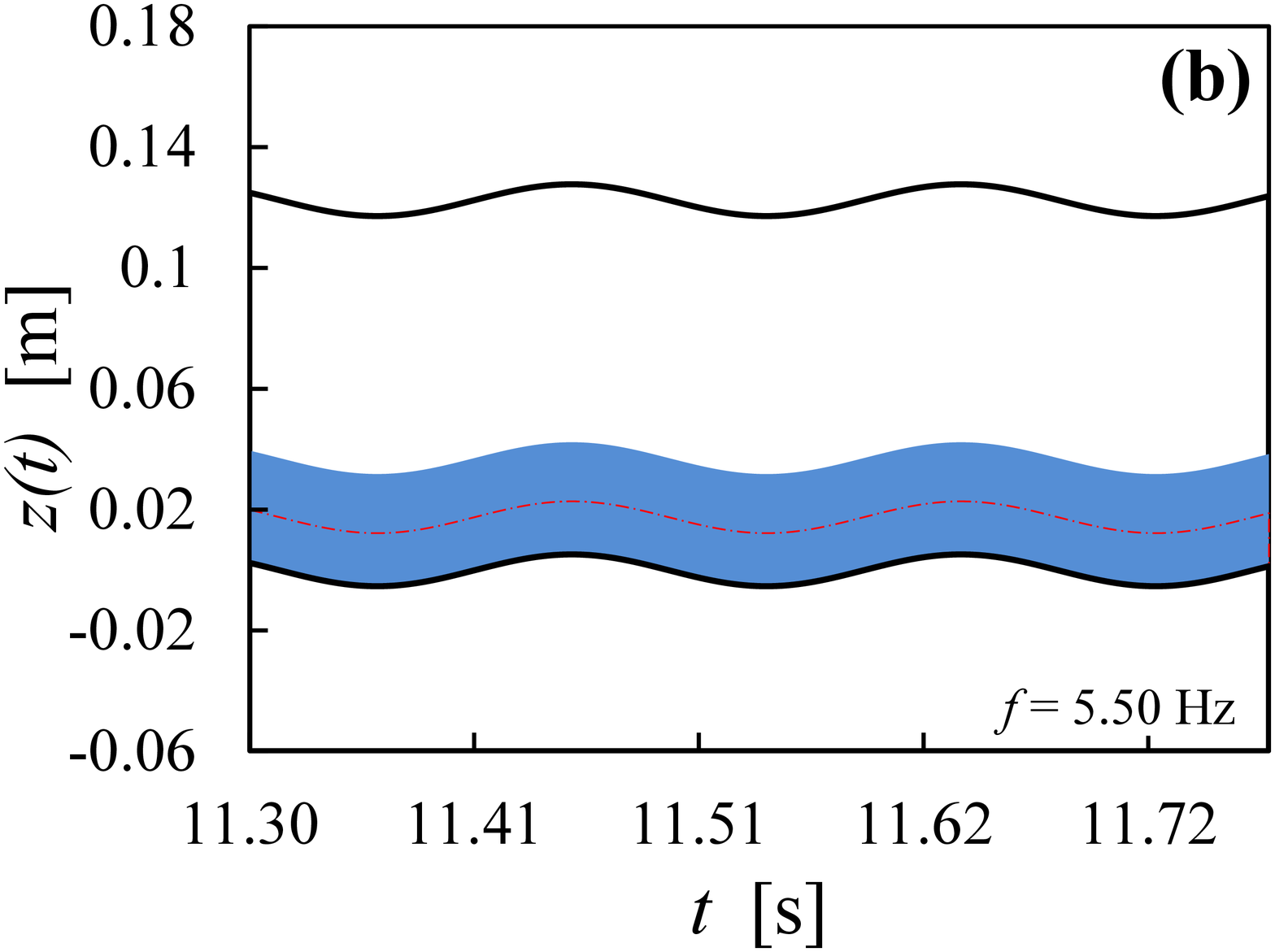}
\includegraphics[width=0.49\columnwidth]{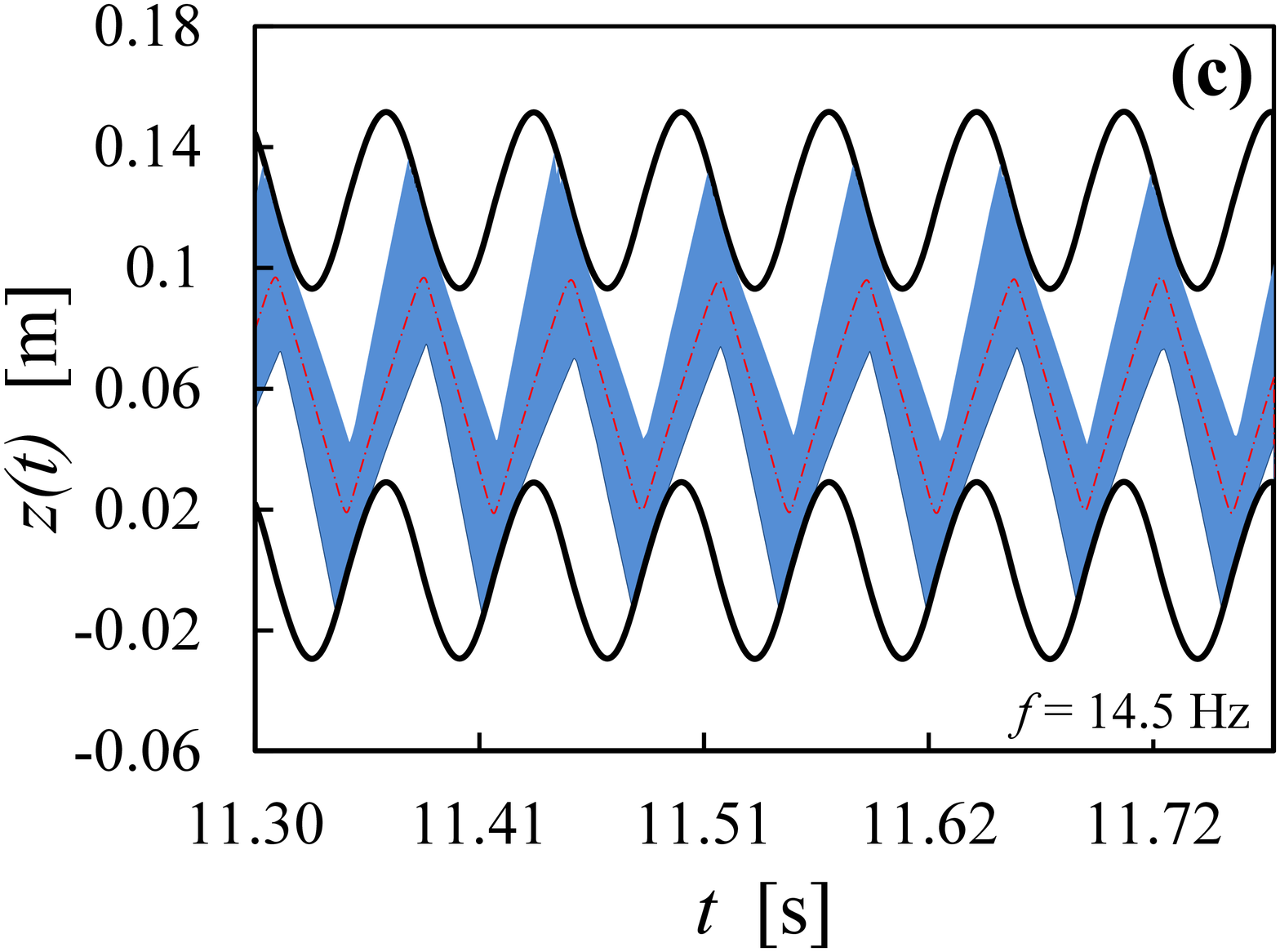}
\includegraphics[width=0.49\columnwidth]{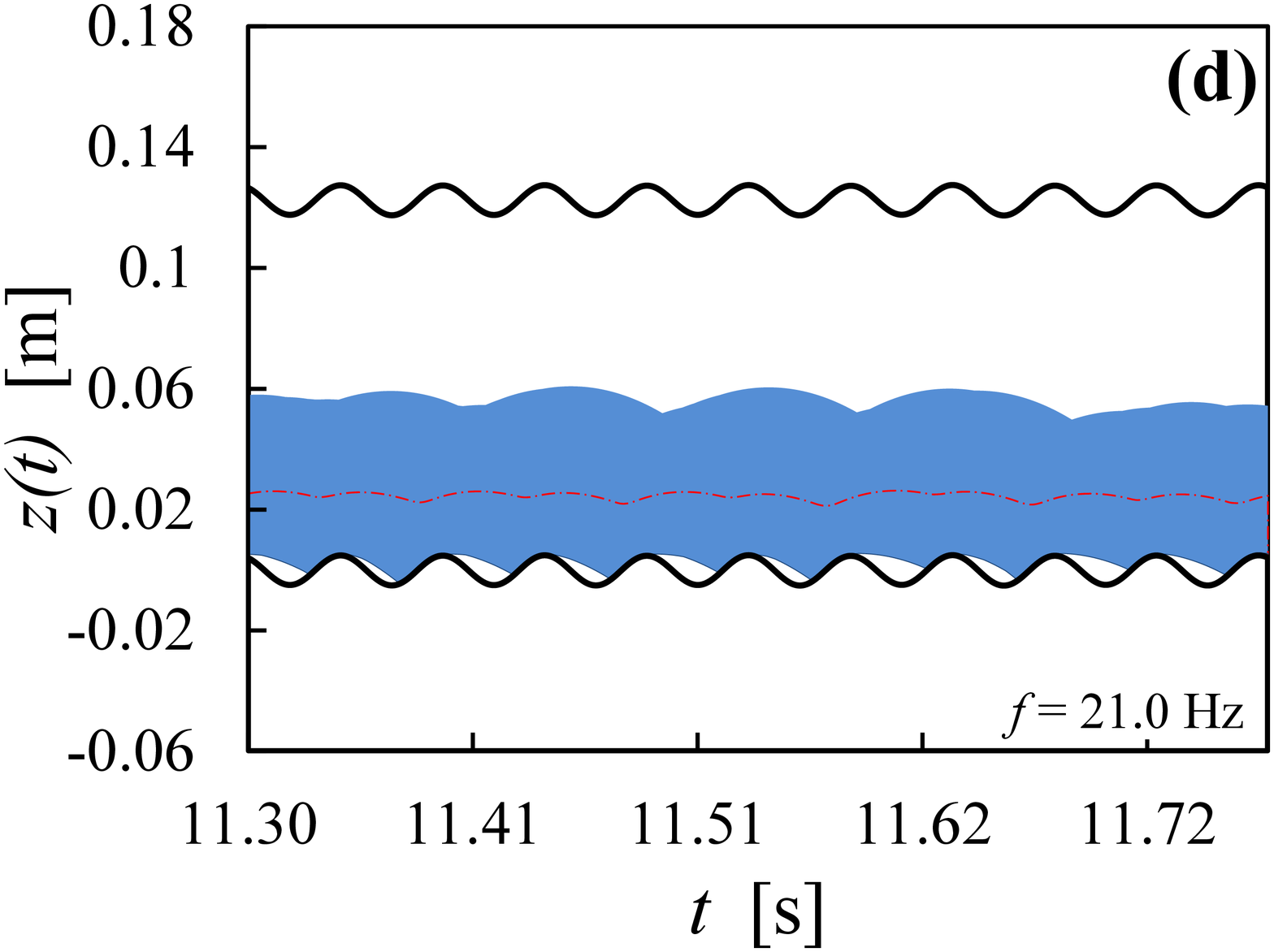}
\caption {(Color online) (a) Frequency response function of the system with $N=250$, $\gamma_n=3660.0$ kgm$^{-1/2}$s$^{-1}$ and $\gamma_t=10980.0$ kgm$^{-1/2}$s$^{-1}$. We plot the maximum amplitude of the oscillation $z_{max}$ of the primary mass as a function of the excitation frequency $f$ imposed to the base for different heights $L_z$ of the enclosure (see legend) and compare with the theoretical FRF for an empty enclosure (black solid line). Lines joining symbols are only to guide the eye. In panels (b), (c) and (d), we plot the trajectory of the enclosure (black solid lines define the floor and ceiling) and the motion of the granular sample inside (colored area defined by the position of the uppermost and lowermost particle at each time; the dotted red line is the position of the center of mass) for the optimum height $L_z=0.12255$m: (b) $f=5.5$Hz, (c) $14.5$Hz and (d) $f=21.0$Hz.}
\label{fig2}
\end{figure}

Figure \ref{fig2}(a) shows the frequency response function (FRF) ---i.e., $z_{max}$ as a function of $f$--- of the system with $N=250$ for a given choice of the granular interaction and three different height $L_z$ of the enclosure in comparison with the response obtained when the enclosure is empty. As discussed in previous studies \cite{Saeki,Papalou,Sanchez1}, there exists an optimum enclosure height (in this case $L_z=0.12255$m) for which the maximum attenuation is achieved. The resonant frequency is shifted due to the added mass (the granular mass), but in a non-trivial way, with overshoot and undershoot effective masses \cite{Sanchez2}. The motion of the granular bed inside the enclosure for the optimum enclosure height can be seen in Fig. \ref{fig2}(b), (c) and (d) for different values of $f$. The grains behave as a more or less dense lumped mass for a wide range of frequencies ($0<f<18$Hz). However, for high frequencies and tall enclosures, the grains enter a gas-like state \cite{Saluena}.

\begin{figure}[t]
\includegraphics[width=0.49\columnwidth]{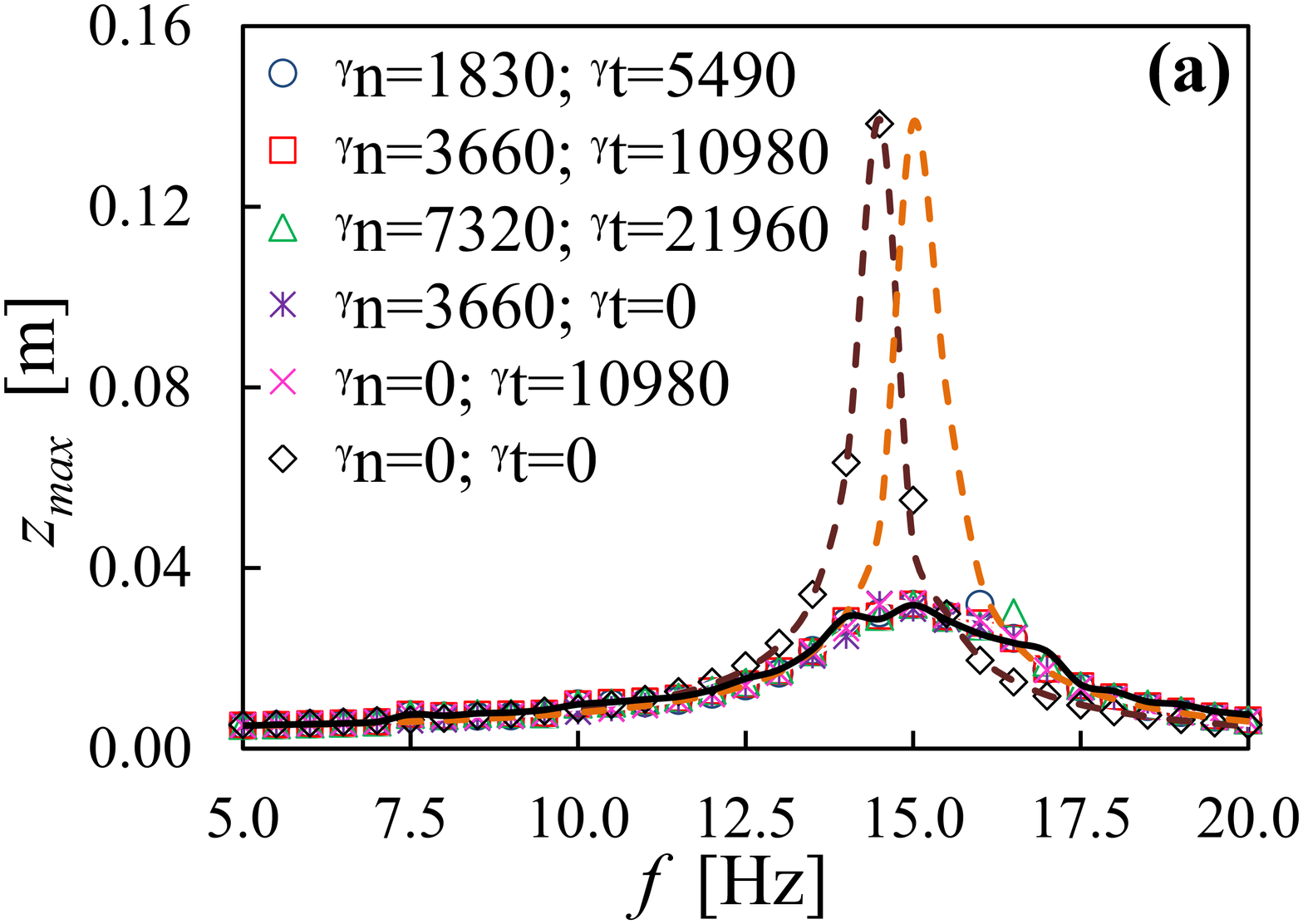}
\includegraphics[width=0.49\columnwidth]{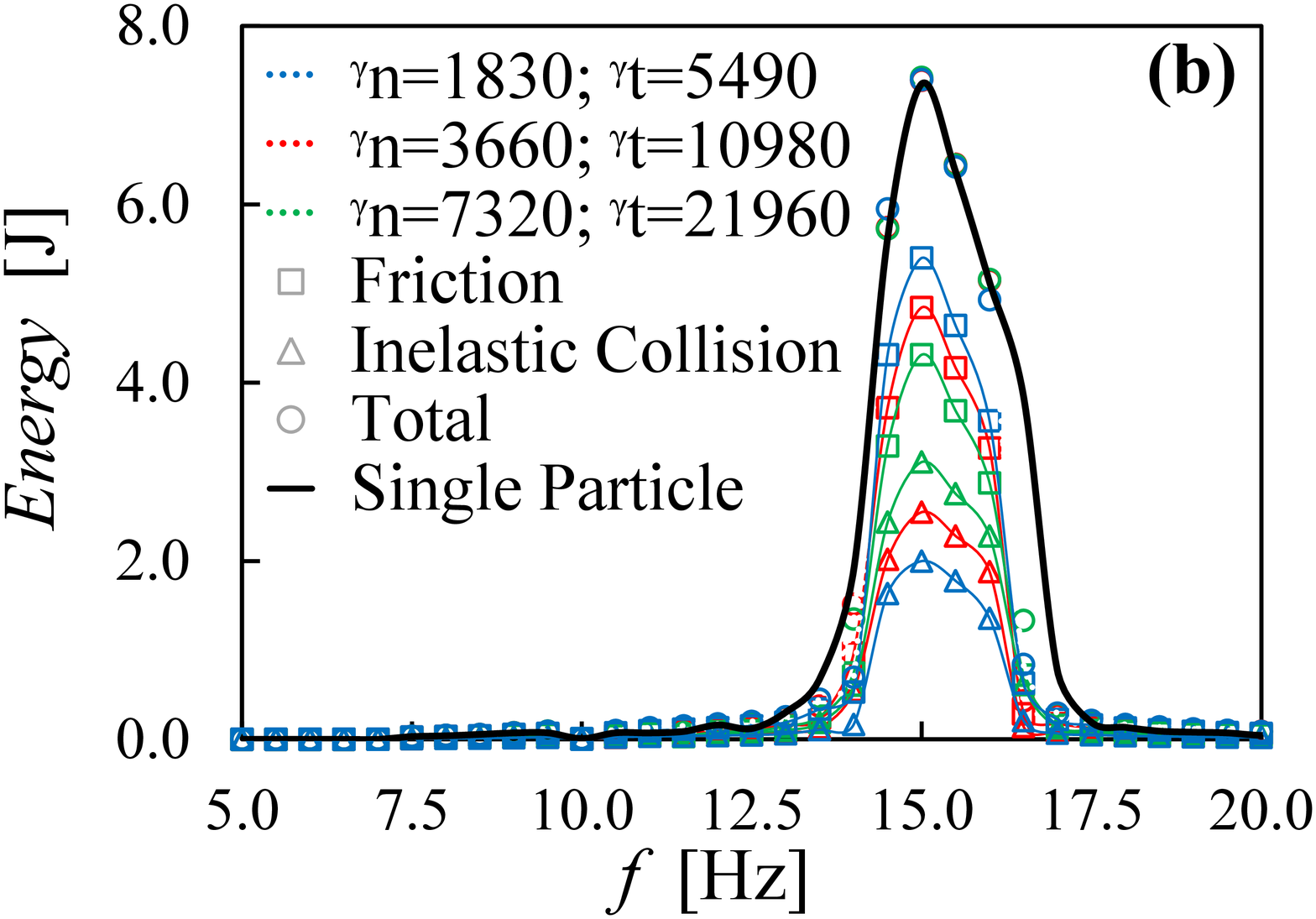}
\includegraphics[width=0.49\columnwidth]{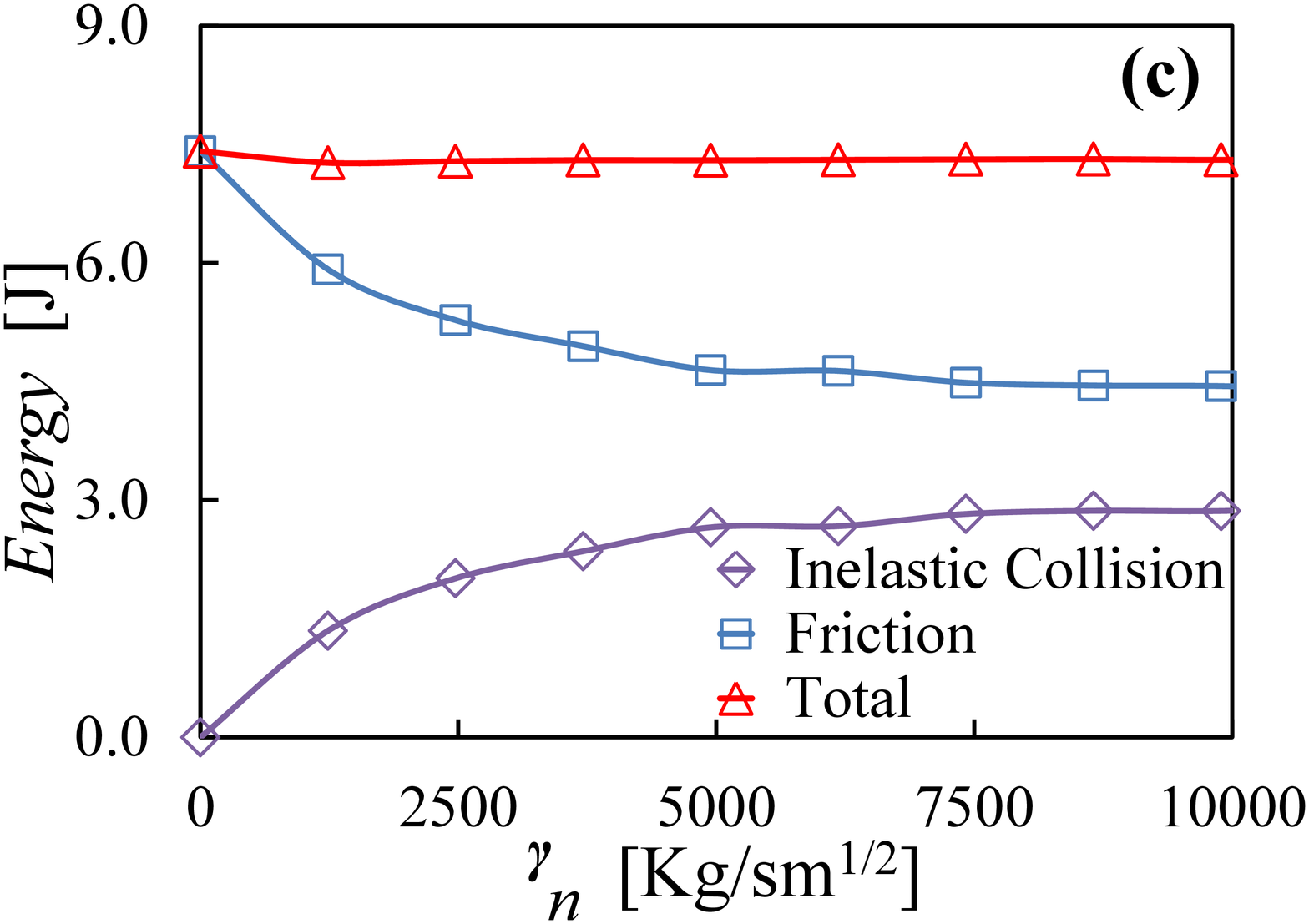}
\includegraphics[width=0.49\columnwidth]{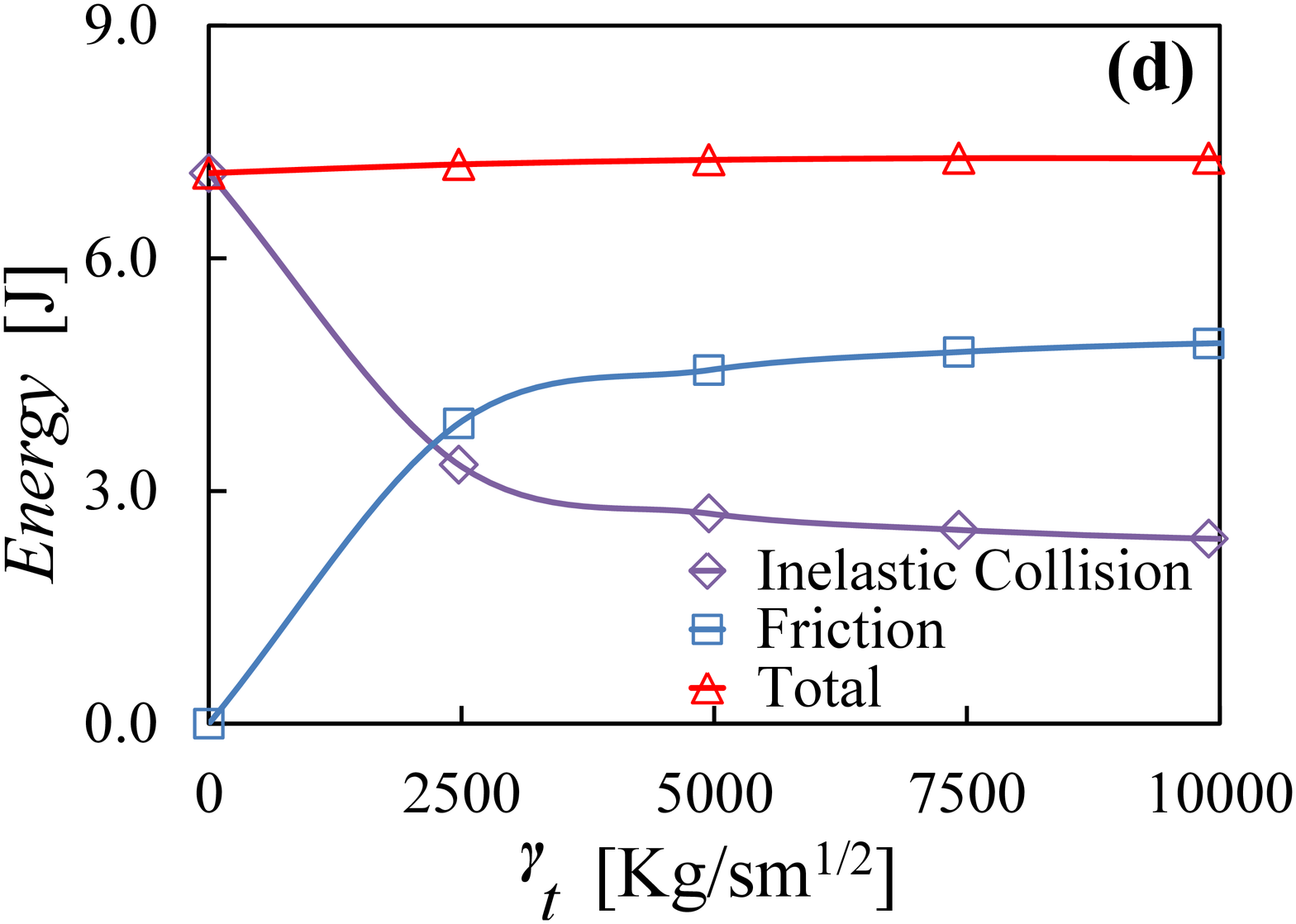}
\caption {(Color online) (a) The FRF for a PD of optimum height ($L_z=0.12255$m) for $N=250$ particles with different interaction parameters $\gamma_n$ and $\gamma_t$ (see legend for the values used in units of kgm$^{-1/2}$s$^{-1}$). The dashed lines correspond to the analytical solution for: an empty enclosure (orange), and an empty enclosure with primary mass corrected to $M+m_p$ (brown). The black solid line corresponds to the zero restitution single particle model. (b) The energy dissipated per cycle as a function of $f$ for the frictional (squares) and collisional (triangles) modes for three different particle--particle interactions (see legend). The total energy dissipated (circles) is independent of $\gamma_n$ and $\gamma_t$. The black solid line corresponds to the zero restitution single particle model. (c) Energy dissipated as a function of the collisional dissipation $\gamma_n$ at the resonant frequency for $\gamma_t=10980.0$ kgm$^{-1/2}$s$^{-1}$. (d) Energy dissipated as a function of the frictional dissipation $\gamma_t$ at the resonant frequency for $\gamma_n=3660.0$ kgm$^{-1/2}$s$^{-1}$.}
\label{fig3}
\end{figure}

\section{Effect of particle--particle interactions}

In order to asses the effect of the particle--particle interactions, we plot in Fig. \ref{fig3}(a) the FRF for the optimum enclosure height for different values of $\gamma_n$ and $\gamma_t$. The effective normal and tangential restitution coefficient is an exponentially decaying function of these parameters and depend on the relative velocity at impact \cite{Schafer}. As we can see, different interaction parameters yield the same FRF, suggesting a universal response. Notice that even eliminating the frictional character of the particles or, alternatively, eliminating the dissipative nature of the normal interaction is not sufficient to induce a change in the FRF over a wide range of frequencies. As it is to be expected, eliminating both, the normal and tangential dissipative part of the interaction converts the system into a conservative molecular-like system which yields no attenuation of the response [see diamonds in Fig. \ref{fig3}(a)]. In this case only a shift of the resonant frequency is observed due to the added mass $m_p$ [see brown dashed line in Fig. \ref{fig3}(a) for the analytical solution]. We have also considered different material properties such as Young modulus and dynamic friction coefficient, but have observed no change in the FRF. 

The energy dissipated in each oscillation cycle is shown in Fig. \ref{fig3}(b) as a function of the excitation frequency. Notice that the frictional dissipation and the collisional dissipation are higher around the resonant frequency. The proportion of energy dissipated through one mode or the other (collisional or frictional) depends on the actual interaction parameters. However, the total energy dissipated is independent of the friction and restitution characteristics of the particles.

To further explore the extent of the apparent universal response of the PD, we have considered a wide range of the dissipative interaction parameters $\gamma_n$ and $\gamma_t$. As it can be seen in Figs. \ref{fig3}(c), where the response of the system with $N=250$ at the resonant frequency is considered, increasing $\gamma_n$ at constant $\gamma_t$ leads to an increase in the energy dissipated though the collisional mode. The converse is true if the frictional parameter $\gamma_t$ is increased [see Fig. \ref{fig3}(d)]. However, the total dissipated energy remains constant even if $\gamma_n$ or $\gamma_t$ drops to zero. Hence, the system is able to dissipate the same amount of energy irrespective of the dissipation modes available to the particles.

\section{Origin of the universal response}

We speculate that this universal response is found whenever a large number of particles is used and the motion of the granular bed is set into a more or less dense lumped mass, as oppose to a gas-like state. For a dense granular layer, the number of collisions per unit time as the bed collides with the boundaries increases dramatically due to an effective inelastic collapse \cite{McNamara,Kadanoff}. Although \emph{inelastic collapse} refers to a mathematical divergence of the number of collisions per unit time when instantaneous interactions are considered, real systems do also exhibit a remarkable increases in the collision rate \cite{Kadanoff,Poschel}. The behavior of such granular systems has been recently proven to be well modeled by a single mass $m$ with an effective zero restitution coefficient and no frictional properties \cite{Chung}. Also, recent simulations with no frictional components where able to fit experimental data on PDs in microgravity \cite{Bannerman}. This fact ---that friction can be neglected--- is evidenced in Fig. \ref{fig3}(d). However, according to Fig. \ref{fig3}(c), restitution can also be set to unity and any non-vanishing friction will suffice to render the universal FRF shown in Fig. \ref{fig3}(a). This indicates that a much wider set of interactions can lead to the effective inelastic collapse than previously shown.

We have solved a simple one-dimensional model where a single particle of mass $m_p$ and zero restitution coefficient moves between the floor an ceiling. The model is adapted from \cite{Friend}. The results are shown in Figs. \ref{fig3}(a) and \ref{fig3}(b) with black solid lines. It is clear that such simple model provides the essence to describe the PD. Hence, design can be based in this simple model without worrying about a careful selection of the particle properties. Previous workers have used slightly more complicated models where the restitution coefficient was used as a fitting parameter \cite{Friend,Bannerman}. Our results indicate that this complication may be unnecessary.

\section{Limits to the universal response}

We now turn into surveying the limits of this universal response. Inelastic collapse is known to take place only if a large number of grains at high densities are considered. Therefore, we can set the system into a regime where this universal response does not apply by reducing $N$ or by promoting a diluted granular state in the enclosure.

\begin{figure}[t]
\includegraphics[width=0.49\columnwidth]{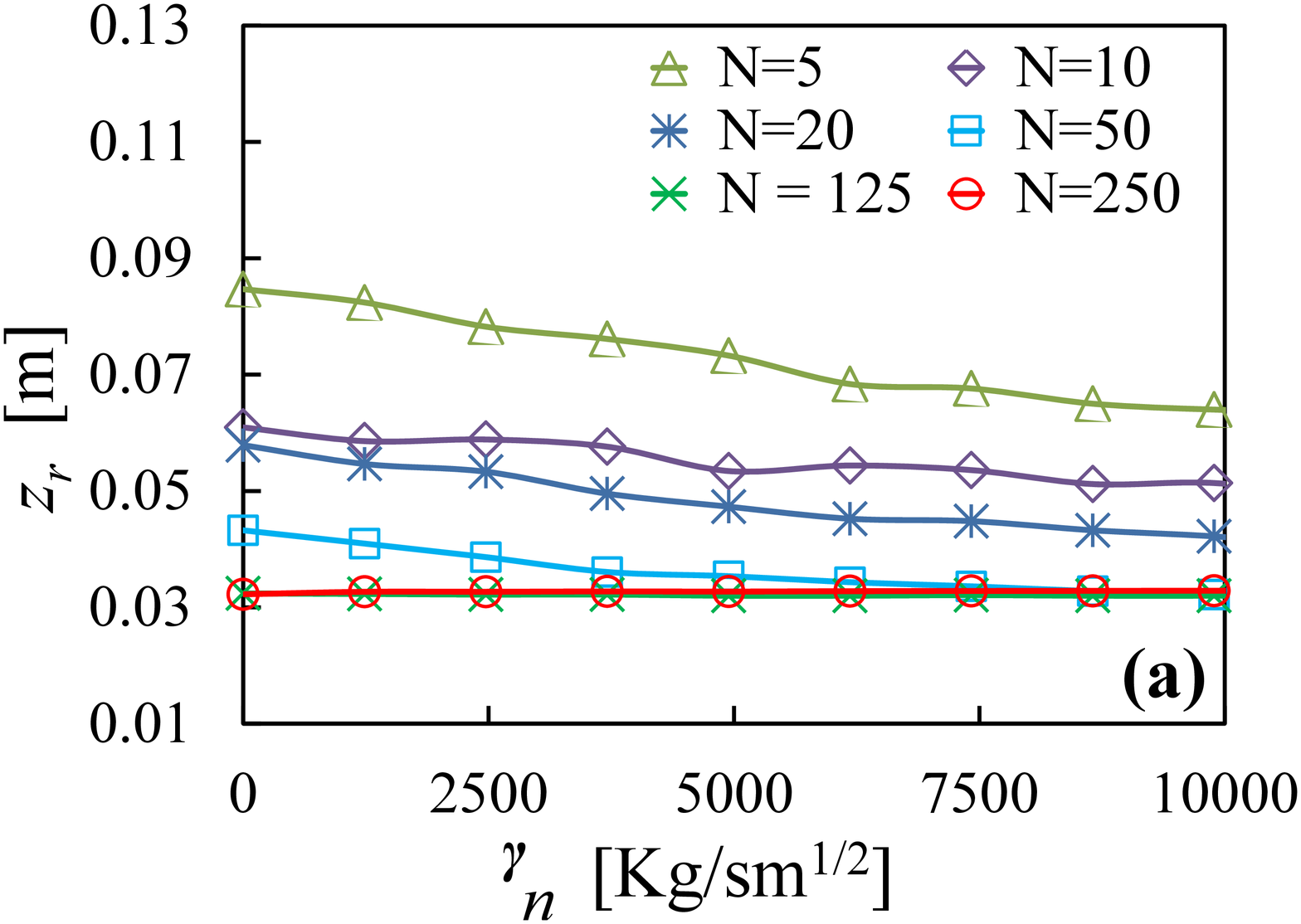}
\includegraphics[width=0.49\columnwidth]{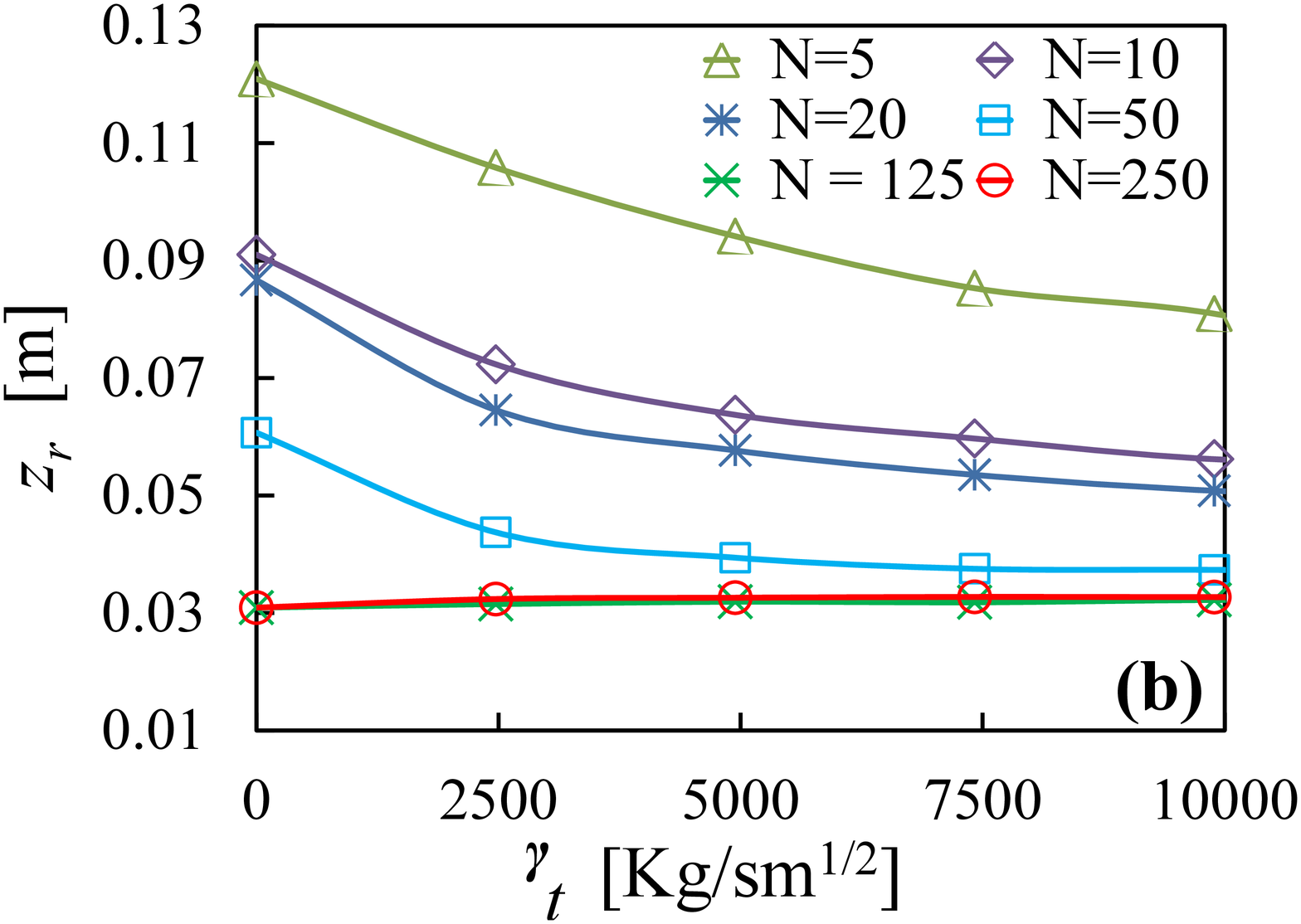}
\caption {(Color online) (a) The vibration amplitude $z_r$ at resonance as a function of the collisional dissipation $\gamma_n$ for $\gamma_t=10980.0$ kgm$^{-1/2}$s$^{-1}$ for different number $N$ of grains in the enclosure (see legend). The particle size is chosen to yield a total particle mass $m_p=0.227$ kg. (b) Same as (a) but the dependence on $\gamma_t$ is considered for $\gamma_n=3660.0$ kgm$^{-1/2}$s$^{-1}$.}
\label{fig4}
\end{figure}

In Fig. \ref{fig4}, we present the variation of the amplitude of vibration $z_{r}$ at the resonant frequency as a function of the normal [Fig. \ref{fig4}(a)] and tangential [Fig. \ref{fig4}(b)] dissipative parameters for different $N$. For these simulations, we have changed the diameter of the particles in order to keep the total mass $m_p$ of the grains constant as $N$ is changed. For $N>100$, a constant (\emph{universal}) response is recovered, whereas smaller systems present a better attenuation as any of the two dissipative properties ($\gamma_n$ or $\gamma_t$) is increased, in accord with intuition. These results confirm the speculation that the universal response only applies if a relatively large number of particles is involved. Of course, the values of $N$ at which this universal response is reached will depend on the horizontal size of the enclosure. We estimate from our simulations that whenever the enclosure is filled with three or more layers of particles the system response near resonance becomes independent of the particle--particle interaction.  

\begin{figure}[t]
\includegraphics[width=0.49\columnwidth]{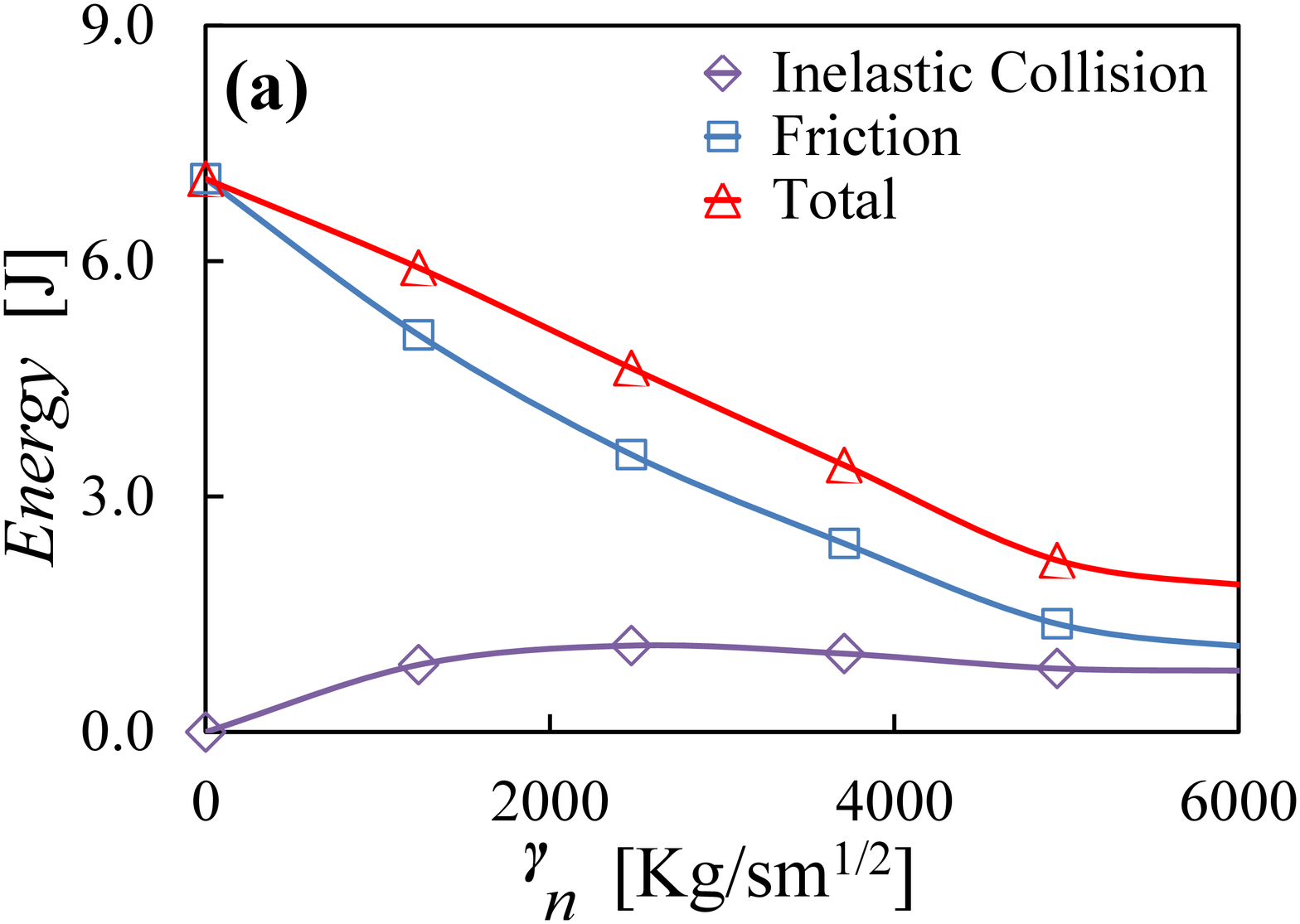}
\includegraphics[width=0.49\columnwidth]{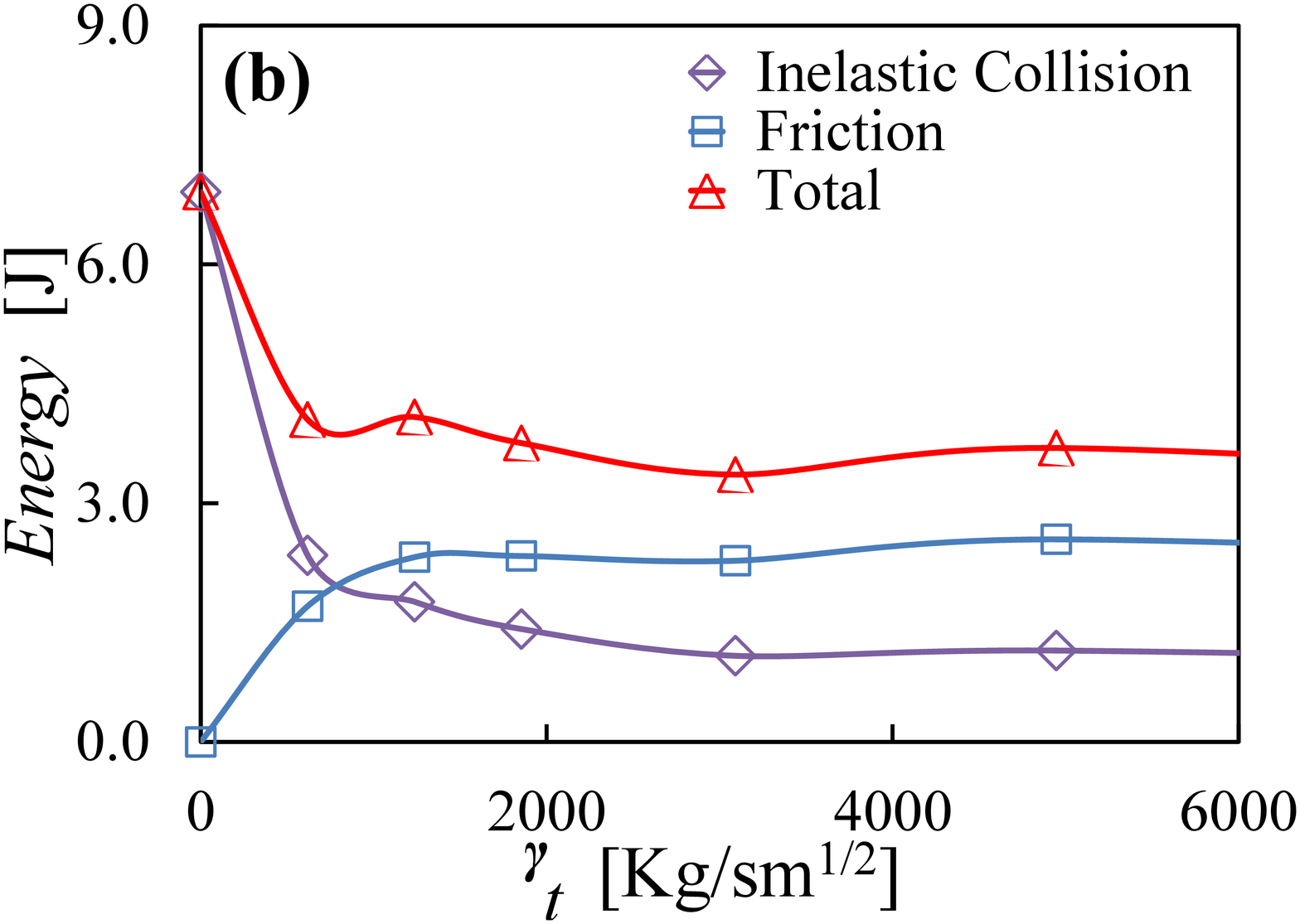}
\caption {(Color online) (a) The energy dissipated per cycle at resonance as a function of the collisional dissipation $\gamma_n$ for a large enclosure ($L_z=0.282$m) with $N=250$ particles and $\gamma_t=10980.0$ kgm$^{-1/2}$s$^{-1}$. (b) Same as (a) but the effect of the frictional dissipation $\gamma_t$ is considered for $\gamma_n=3660.0$ kgm$^{-1/2}$s$^{-1}$.}
\label{fig5}
\end{figure}

A way to induce a gas-like behavior of the granular sample in the enclosure ---so as to create a dilute regime where an effective inelastic collapse is not expected--- is to increase the height of the cavity. In Fig. \ref{fig5}(a) and (b) we plot the energy dissipated per cycle as a function of $\gamma_n$ and $\gamma_t$ for an enclosure with $L_z=0.282$m. For this large $L_z$, the granular sample expands significantly and does not move as a lumped mass, so reducing dramatically the number of collisions per unit time. The final results is an effective dissipation that depends on the particle--particle dissipative interaction. Interestingly, in this regime, increasing $\gamma_n$ or $\gamma_t$ leads to a decrease of the total energy dissipated. Since PDs are designed to optimize attenuation in a number of applications, the size of the enclosure generally promotes the dense lumped mass motion of the grains inside \cite{Simonian}. Therefore, in many working conditions of interest, the system will be in a regime where the universal FRF is obtained.

\section{Conclusions}

We have shown that a basic phenomenon (inelastic collapse) leads to a universal response of a PD ---in the sense that the particle--particle interaction becomes irrelevant. This allowed us to determine the limits of this universality: relatively large numbers of particles must be used and the system has to be set in a state of dense lumped mass. 

It is worth mentioning that the particular contact force model used for the simulations is of little relevance to the results presented here. We have shown that a much simplistic model (the zero-restitution-single-particle model) also shows the same response. Indeed, the proposed universality implies that details of the interactions are not relevant. Moreover, despite our studies being done on a system vibrating in the direction of gravity, we expect results will apply to horizontally vibrating PDs.

This universal response is consistent with some observations in experiments and simulations where a few values of the material properties where tested \cite{Marhadi,Chen,Bai,Lu, Duan} and can explain the unexpected agreement between simplified models and complex experiments \cite{Bannerman}. It is worth mentioning that powders, as opposed to granulars, may not follow this universal response even at resonance. Powders are fine graded particles and the effects of the hydrodynamics of the surrounding air affects the motion of the particles to a large extent. Fine powders will expand due to air-particle interactions and the inelastic collapse will be unlikely. Some preliminary experiments with PDs using fine powders seem to confirm this \cite{Marhadi}.

The suitability of PDs to work in harsh environments can be understood as a consequence of this phenomenon. Extreme temperatures and pressures may induce mild changes in frictional properties, but these will not alter the PD response. More importantly, degradation of the particles during operation due to wear, deformation and fragmentation are not likely to compromise the PD performance. Changes in friction or restitution are unimportant. Although we have not studied particles of different shapes, we speculate that fragments of particles may be as effective as the original particles as long as they are not fine graded. Moreover, fragmentation can only increase $N$, which should not take the system out of the universal FRF. This is the underlying phenomenon that explains the characteristic low maintenance required for this devices. Notice however that very high temperatures may weld particles together inducing an effective reduction of $N$ which can reduce vibration attenuation.

Design of PDs can be greatly simplified by choosing to work with large $N$ and using a simple model such as the zero restitution single mass used here. Under these conditions, the selection of the particle material properties is unimportant for the PD performance and one can focus, for example, on cost effectiveness. 

\section*{Acknowledgements}
LAP acknowledges support from CONICET (Argentina).

\bibliographystyle{elsarticle-num}

\bigskip 

\Large
Figure Captions
\normalsize

Figure 1: (Color online) Schematic view of a particle damper. ($M$) the primary mass of the structure, ($m_p$) the total mass of the particles in the enclosure, ($K$) the spring constant, ($C$) the viscous structural damping, ($B$) the vibrating base where the displacement is imposed.

Figure 2: (Color online) (a) Frequency response function of the system with $N=250$, $\gamma_n=3660.0$ kgm$^{-1/2}$s$^{-1}$ and $\gamma_t=10980.0$ kgm$^{-1/2}$s$^{-1}$. We plot the maximum amplitude of the oscillation $z_{max}$ of the primary mass as a function of the excitation frequency $f$ imposed to the base for different heights $L_z$ of the enclosure (see legend) and compare with the theoretical FRF for an empty enclosure (black solid line). Lines joining symbols are only to guide the eye. In panels (b), (c) and (d), we plot the trajectory of the enclosure (black solid lines define the floor and ceiling) and the motion of the granular sample inside (colored area defined by the position of the uppermost and lowermost particle at each time; the dotted red line is the position of the center of mass) for the optimum height $L_z=0.12255$m: (b) $f=5.5$Hz, (c) $14.5$Hz and (d) $f=21.0$Hz.

Figure 3: (Color online) (a) The FRF for a PD of optimum height ($L_z=0.12255$m) for $N=250$ particles with different interaction parameters $\gamma_n$ and $\gamma_t$ (see legend for the values used in units of kgm$^{-1/2}$s$^{-1}$). The dashed lines correspond to the analytical solution for: an empty enclosure (orange), and an empty enclosure with primary mass corrected to $M+m_p$ (brown). The black solid line corresponds to the zero restitution single particle model. (b) The energy dissipated per cycle as a function of $f$ for the frictional (squares) and collisional (triangles) modes for three different particle--particle interactions (see legend). The total energy dissipated (circles) is independent of $\gamma_n$ and $\gamma_t$. The black solid line corresponds to the zero restitution single particle model. (c) Energy dissipated as a function of the collisional dissipation $\gamma_n$ at the resonant frequency for $\gamma_t=10980.0$ kgm$^{-1/2}$s$^{-1}$. (d) Energy dissipated as a function of the frictional dissipation $\gamma_t$ at the resonant frequency for $\gamma_n=3660.0$ kgm$^{-1/2}$s$^{-1}$.

Figure 4: (Color online) (a) The vibration amplitude $z_r$ at resonance as a function of the collisional dissipation $\gamma_n$ for $\gamma_t=10980.0$ kgm$^{-1/2}$s$^{-1}$ for different number $N$ of grains in the enclosure (see legend). The particle size is chosen to yield a total particle mass $m_p=0.227$ kg. (b) Same as (a) but the dependence on $\gamma_t$ is considered for $\gamma_n=3660.0$ kgm$^{-1/2}$s$^{-1}$.

Figure 5: (Color online) (a) The energy dissipated per cycle at resonance as a function of the collisional dissipation $\gamma_n$ for a large enclosure ($L_z=0.282$m) with $N=250$ particles and $\gamma_t=10980.0$ kgm$^{-1/2}$s$^{-1}$. (b) Same as (a) but the effect of the frictional dissipation $\gamma_t$ is considered for $\gamma_n=3660.0$ kgm$^{-1/2}$s$^{-1}$.

\end{document}